\documentclass[12pt]{article}

\usepackage{vmargin}   
\setmarginsrb{3cm}{2cm}{3cm}{2cm}{1cm}{1cm}{1cm}{1cm}   
\usepackage{amsmath}   
\usepackage{amssymb}  
\usepackage[dvips]{graphicx}  
\usepackage{rotating}   
\usepackage{psfrag} 
\usepackage{fancyhdr}

\author{Karim Noui \\
Laboratoire de Math\'ematiques et de Physique Th\'eorique \\
UMR/CNRS 6083, F\'ed\'eration Denis Poisson \\
Facult\'e des Sciences et Techniques\\
Parc de Grandmont, 37200 Tours }
\title{{\bf Motion in Quantum Gravity}}
\date{} 

\begin{document} 
\sloppy
\maketitle

\abstract{
We tackle the question of motion in Quantum Gravity: what does motion mean
at the Planck scale? Although we are still far from a complete answer   
we consider here a toy model in which the problem can be formulated and
resolved precisely. The setting of the
toy model is three dimensional Euclidean gravity. Before studying the model in detail,
we argue that Loop Quantum Gravity may provide a very useful approach when discussing the question
of motion in Quantum Gravity.
}

\section{Introduction}
\subsection{The problem of defining motion in quantum gravity}
Motion is fundamentally a classical notion: ``it refers to a change of position in space".
When we talk about quantum physics or relativity, the definition of motion 
has to be made more precise, for either the notion of position is not well-defined (in quantum physics)
or the notion of space-time has to be rethought (in relativity). Indeed, when we turn on the Planck constant
$\hbar$, matter is described in terms of wave functions which are not localized, so a point particle can, a priori, be everywhere
at any time; one needs to introduce coherent states, for instance, to recover the reassuring 
notion of trajectory at the classical limit. When we turn on the light speed $c$, time is no longer absolute but becomes intimately mixed with 
spatial coordinates;
we need to make precise what time means if we are to define the motion properly. When we turn on $\hbar$
and $c$ together, matter fields and their interactions are beautifully described within the Quantum
Field Theory framework, which is rather non-intuitive, but one has to work quite hard to make a bridge to the classical world. 

What happens if we now introduce the gravitational constant $G$ into the scenario? Newton gave us laws
to explain the attraction between massive bodies and created tools for studying their trajectories,
as long as the bodies are not too ``small" and their velocities not too ``high".
Turning on $c$ and $G$ together leads to general relativity, where space-time becomes a 
dynamical entity which interacts deeply with all types of matter and energy; we definitively lose the classical 
absolute background that is so necessary for defining the classical notion of motion. It is, all the same, possible
to extend this notion and adapt it to general relativity; the geodesics for instance correspond to trajectories
of infinitely light particles evolving in a space-time with which we assume they do not interact. The determination
of trajectories without neglecting the self-force is a much more subtle, but more realistic and interesting, problem \cite{barack}.
In particular, it puts forward the trivial but fundamental fact that the point-like description for massive matter-fields
is completely meaningless in general relativity because it leads to black hole singularities. Thus, to have a proper 
description of motion in general relativity, one needs to consider extended matter fields, which obviously makes the problem
much more complicated at the technical and conceptual levels. 

Defining motion in a theory where all the fundamental constants $\hbar$, $G$ and $c$ are switched on 
is clearly too ambitious a problem. It is certainly too early to investigate it and  
we do not claim to solve it here.  Rather, we would like to raise the preliminary questions that 
naturally arise while addressing such a concept, and to see if it is possible to answer some 
of them precisely.
It goes without saying that the question of the fundamental structure of space-time comes first to 
mind. Even though it is commonly believed that space-time is no longer described in terms of a differential
manifold at the Planck scale, it is also honest to claim that no one knows precisely how space-time appears 
at this scale. Nonetheless, there exist very fascinating proposals  
that one can take seriously when investigating the question of motion in quantum gravity. 

\subsection{Quantum gravity}
It is indeed openly recognized that a complete and consistent quantisation of gravity that would 
give a precise description of
space-time at the Planck scale is still missing. Many ways to attack this problem have been explored over 
the last twenty years, the two most popular surely being String Theory \cite{string} and Loop Quantum Gravity \cite{lqg}.
While both these approaches aim to understand the deep and fundamental structure of space-time, they have
developed very different strategies and achieved, so far, rather distinct results. For instance, String Theory
proposes a version of quantum space-time with extra-dimensions whereas, in Loop Quantum Gravity, space-time
is fundamentally four dimensional with three dimensional space slices which are discrete in some precise sense.
The discreteness of space is, in fact, one of the most beautiful but intriguing achievements of Loop Quantum
Gravity. Even if this result is controversial and unconfirmed, it makes Loop Quantum Gravity quite a fascinating
approach that certainly deserves to be investigated, to at least understand how far it can bring us towards the
Planck regime. 

To achieve this discreteness, Loop Quantum Gravity has  adopted a very ``conservative" point of view,
namely the canonical quantization of Einstein-Hilbert theory reformulated in terms of Ashtekar variables \cite{ashtekar}
with no extra-fields or extra-dimensions: only gravity and the laws of quantum physics.
The basic idea is therefore very simple. One could naturally asks why such a simple idea has not been explored until
recently, for gravity and quantum physics have existed for almost a century. In actuality, 
quantizing general relativity with the "standard" tools of quantum mechanics has been investigated from its inception, but it immediately faced huge problems: the canonical quantization \`a la ADM \cite{ADM}
leads to a system of highly non-linear equations (the famous constraints) which are simply impossible to solve whereas
the perturbative path integral quantization makes no sense since gravity is non-renormalizable.

Does Loop Quantum Gravity overcome these fundamental difficulties? A honest answer would be: we still do not know.
Why? Because, so far, Loop Quantum Gravity has ``only" opened a new route towards the quantization of gravity, and
we are still far from the end of the story. Nonetheless, the road is very fascinating.  Among other things, it has allowed us to introduce
very interesting new ideas, such as (so-called) ``background independence", and to formulate, for the first time, questions about the structure of space-time
at the Planck scale, in a mathematically well-defined way. Loop Quantum Gravity is not (yet)
a consistent theory of quantum gravity, but it has proposed very exciting preliminary results.

The starting point has been the discovery by Ashtekar of a new formulation of gravity.
In the Ashtekar variables, gravity reveals strong similarities with 
$SU(2)$ Yang-Mills theory and, when starting to quantize
general relativity, one makes use of the techniques developed for gauge theories. In particular, the physical states of quantum gravity are expected to be constructed from so-called
spin-network states, which are a generalization of the Wilson loops and are 
associated to ``colored three dimensional topological graphs". Thus, space slices are
described in terms of graphs at the Planck regime and their geometrical content is encoded into the coloration of each
graph. 
Roughly, colored graphs are for quantum gravity what the quantum numbers $(n,\ell,m)$ are
for the hydrogen atom: $(n,\ell,m)$ characterize states of the electron in the hydrogen atom
and a colored graph characterizes a state of quantum geometry.
Spin-network states are shown to be eigenstates of certain geometrical operators, such as the area and the volume operators,
with discrete eigenvalues, making quantum spaces discrete in Loop Quantum Gravity.
The theoretical framework for describing these quantum geometries is mathematically very well-defined and 
has already been exposed in several reference books and articles \cite{lqg}.

If we choose to view Loop Quantum Gravity as a starting point for understanding motion at the Planck
scale, there comes the question of the description of the matter fields, and of their coupling to quantum gravity.
Contrary to String Theory, Loop Quantum Gravity is, a priori, a quantization of pure gravity. A way to include matter
in that scheme consists in first considering the classical coupling between the Einstein-Hilbert action with a (Klein-Gordon, 
Dirac, Maxwell or Yang-Mills)  field and then quantizing the coupled system. 
Prior to quantization, one has to reformulate the coupled theory in terms of Ashtekar variables, which is in fact immediate.
Thus, not only it is, in principle, possible to consider all the matter fields of the standard model but also one can directly include
super-symmetry in that scheme.

As one could expect, in general, the presence of 
extra fields makes Loop Quantum Gravity much more complicated, but it has been shown that Loop Quantum
Gravity techniques can be extended to these cases, and quantum gravity effects  
make the resulting Quantum Field Theories
free of UV and IR divergences \cite{thiemann-matter}. Thus, one concretely realises the old idea that UV
divergences in Quantum Field Theory are a reflection of our poor understanding of the physics at very short distances
and quantum gravity should provide a regulator
for Quantum Field Theory. However, we do not know how to solve the dynamics explicitly, that is, we do not have any ideas
for the solution of the quantum equations of motion. 

One idea for overcoming this difficulty is  
to assume that the matter field 
would be so ``light" that it would not affect the (quantum) space-time structure and would
follow the quantum analogue of a geodesic curve. This hypothesis appears immediately inconsistent, because there exists 
no regime in which space-time is quantized and the matter coupling to gravity can be neglected.
A quantum gravity phenomenology has been developed  to provide a more or less realistic picture of the effects
of the quantized background on the motion. In that framework, many have predicted, for instance, a violation of Lorentz
invariance, which manifests itself in the dispersion relation of some particles. These results have been discussed and
criticized extensively in the literature. We will not continue this discussion here, but we do want to at least underline the fact that
the discreteness of space is the one link that exists between this phenomenology and Loop Quantum Gravity.
It is definitively clear that no one yet has a precise idea of what is motion in Loop Quantum Gravity. 

\subsection{Three dimensional quantum gravity is a fruitful toy model}

One way to be more precise is to study simplified models of quantum gravity. Three dimensional quantum
gravity is such a toy model which has been explored considerably over the last twenty years, starting 
from the fundamental article of Witten who established an amazing relation between 3D quantum gravity and the
Jones polynomials \cite{witten1}. Previously, 3D gravity was supposed to be too trivial to deserve any attention: there are no
local degrees of freedom, there is no gravitational attraction between massive particles --- whose coupling to gravity
creates ``only" a conical singularity in the space-time at the location of the particle. 
This apparent simplicity hides not only incredibly rich mathematical
structures but also a real physical interest in 3D gravity, which may help us understand important conceptual issues
concerning the problem of time, and how to deal with invariance under diffeomorphisms, for instance. 
The discovery of black holes in 3D Lorentzian anti-De Sitter gravity has also greatly increased the 
interest in such a toy model \cite{btz}. 
Many quantization schemes have been
developed, as evidenced in the book  by Carlip\cite{carlip}. The coupling to massive and spinning particles has also been thoroughly studied
at both the classical and quantum levels, and has revealed a close relationship between particle dynamics 
and knot invariants
in three dimensional manifolds \cite{witten1}. 

Naively, it might seem to make no physical sense to quantize gravity coupled with 
point particles: in the regime where space-time becomes quantized, we expect the matter field to be quantized 
as well and then to be described in terms of fields instead of particles. In fact, the coupling to point particles
is not completely devoid of physical interest because it appears to be a good starting point for understanding
the coupling of quantum gravity to quantum fields. 
The first reason is that point particles do exist in 3D general relativity contrary to the four dimensional case.
The second reason is  simply to notice that, if we do not know how to
quantize gravity coupled to matter fields starting from the quantization of the matter fields in a given (flat)
background and then perturbativly quantizing the geometry, we could try the other way around. Indeed, why not first try
quantizing gravity non-perturbativly, keeping the matter classical, and then proceed to the quantization of
the matter degrees of freedom in the quantum background? This point of view makes some sense, as pure quantum gravity is
very well understood in three dimensions. Furthermore, it was very fruitful and lead to the very first full quantization
of a massive self-gravitating scalar field in the context of Euclidean Loop Quantum Gravity \cite{fl,moi1}.

The most important consequence of this study is certainly the fact that quantum gravity turns classical
differential manifolds into non-commutative spaces where the non-commutativity is encoded into the Planck length 
$\ell_P=\sqrt{G\hbar/c^3}$. More precisely, it has been  argued that
a quantum scalar field coupled to Euclidean three dimensional gravity is 
equivalent to a sole quantum scalar field living in a non-dynamical but non-commutative space. 
The emerging non-commutative space appears to be a deformation of the standard 3D Euclidean space
and admits a quantum group, known as the Drinfeld (quantum) double $DSU(2)$, as ``isometry group" \cite{jihad}.
As a consequence, the question of motion in 3D quantum gravity turns into the question of motion
in a non-commutative space. This problem is mathematically very well-defined and admits a precise solution.
In particular, the quantum space admits a fuzzy space formulation and a massive scalar field is described in terms
of complex matrices. Equations of motion are finite difference equations involving the matrix coefficients and their
solutions allows us to understand how the notion of motion is modified in quantum gravity. 
Once again, 3D gravity appears as an incredibly good toy model to have for a first view of fundamental issues.
This article is mainly devoted to explain how motion can be described in 3D Euclidean quantum gravity.

\subsection{Outline of the article}

This article is structured as follows. We start, in Section 1, with a very brief review of Loop Quantum Gravity: 
we focus on the aspects we think are the most important; details can be found 
in numerous informative references \cite{lqg}. We present the main lines of the quantization strategy, then describe
the states of quantum geometry in terms of spin-networks and finally explain in which sense quantum geometries are
discrete, presenting a computation of the spectrum of areas operators.  We also mention open issues concerning the problem
of dynamics: how do we find solutions of the Hamiltonian constraint? 

Section 2 is devoted to giving a precise answer to the question of motion in 3D Euclidean quantum gravity.
It is mainly based on the paper \cite{moi2}.
First, we explain why quantum gravity makes space-time non-commutative in that context. The
emerging non-commutative geometry is a deformation of the classical 3D Euclidean space whose ``isometry"
algebra is a deformation of the Euclidean symmetry algebra as well. 
We describe this non-commutative space and propose different equivalent formulations:
of particular interest is its fuzzy space formulation where it appears as an union of concentric fuzzy spheres. 
Then, we show how to describe the dynamics
of a massive scalar quantum field in such a non-commutative geometry: to be well-defined, the scalar field must
have different a priori independent components; its dynamics is governed by an action very similar to the classical one
but non-local; equations of motion can be written as finite difference equations which couple in general the
different components of the field. We give the solutions when the field is free. When the field is not free, equations
of motion do not admit generically explicit solutions. Faced with this technical difficulty, we perform a symmetry reduction
to simplify the problem and propose a perturbative solution of the reduced system. The solution is interpreted as
the motion of a particle in Euclidean quantum gravity. 

We finish the paper with a Section that contains our conclusion and outlook.

\section{Casting an eye over Loop Quantum Gravity}
Loop Quantum Gravity is a particularly intriguing candidate for a background independent non-perturbative Hamiltonian quantization
of General Relativity. It is based on the Ashtekar formulation of gravity \cite{ashtekar} which is (in a nutshell) a first order formulation where
the fundamental variables are an $SU(2)$ connection $A$ and its canonical variable, the electric field $E$. 

\subsection{The classical theory: main ingredients}
The starting point is the classical canonical analysis of the Ashtekar formulation of gravity.
In this framework, space-time is supposed to be (at least locally) of the form $\Sigma \times \mathbb R$ in order for the canonical theory to 
be well-defined, in particular, for the Cauchy problem to be well-posed.
In terms of these variables, gravity offers interesting similarities with $SU(2)$ Yang-Mills theory that one can exploit
to start quantizing the theory. The connection $A$ is, strictly speaking, the analogue of the Yang-Mills gauge field. 
At this stage of our very brief description of the theory, let us emphasize some aspects which are important for a good understanding of the hypotheses underlying the
construction of Loop Quantum Gravity:
\begin{enumerate}
\item {\it The question of the covariance.} Due to the choice of a splitting $\Sigma \times \mathbb R$ of the space-time manifold,
one is manifestly  breaking the covariance of general relativity! It is the price to pay if one formulates a canonical description of General
Relativity. In standard Quantum Field Theories (QFT), this aspect is not problematical, even if we make an explicit choice of a preferred time, 
because one recovers at the end of the quantization that the Quantum Theory is invariant under the Poincar\'e group. In General Relativity,
the situation is more subtle because making a preferred time choice breaks a local symmetry whereas the Poincar\'e symmetry is a global one
in standard QFT. The consequences of such a choice might be important in an eventual quantum theory of General Relativity. Spin-Foam models
(Section \ref{dynamics}) are introduced partly to circumvent this problem.
\item {\it Where does the group $SU(2)$ come from?} To answer this question, we briefly recall the construction of Ashtekar variables. 
The starting point is the first order formulation of Einstein-Hilbert action \`a la Palatini where the 
metric variables (described in terms of tetrads $e$) and the connection $\omega$ are considered as independent variables:
\begin{eqnarray}
S[e,\omega] \; =\frac{1}{8\pi G} \; \int_{\cal M} \langle e \wedge e \wedge \star F(\omega) \rangle,
\end{eqnarray}
where $\langle,\rangle$ holds for the trace in the fundamental representation of $sl(2,\mathbb C)$
and $\star$ is the hodge map in $sl(2,\mathbb C)$. It becomes clear that the Palatini theory admits the Lorentz group $SL(2,\mathbb C)$ as a local symmetry group.  
Then one performs a gauge fixing, known as the time gauge, which breaks the $SL(2,\mathbb C)$ group into $SU(2)$, its subgroup of rotations.
This is the origin of the symmetry group $SU(2)$ in Loop Quantum Gravity. 
\item {\it The Barbero-Immirzi ambiguity.} In fact, there is a one parameter family of actions which are classically
equivalent to the Palatini action. This remark has been observed first in the Hamiltonian context \cite{Barbero}
before Holst \cite{Holst} wrote the explicit form of the action:
\begin{eqnarray}
S[e,\omega] = \frac{1}{8\pi G} \; \int_{\cal M} 
\left(\langle e \wedge e \wedge \star F(\omega) \rangle \, - \, 
\frac{1}{\gamma} \langle e \wedge e \wedge  F(\omega) \rangle \right).
\end{eqnarray}
$\gamma$ is the Barbero-Immirzi parameter.  The canonical analysis of the Holst action leads to a set of canonical variables
which are a connection $A\equiv \frac{1}{2}(\omega - \gamma^{-1} \star\omega)$ and its conjugated variable $E$.
The variable $A$ is precisely the Ashtekar-Barbero connection. Historically,  Ashtekar found this connection for $\gamma=i$:
he noticed that the expression of the constraints of gravity simplify magically in that context but he had to deal with the so-called
reality constraints to recover the real theory. So far, no one knows how to solve the reality constraint in the quantum theory.
For that reason, the discovery of the Ashtekar-Barbero variable appeared as a breakthrough, for the variables are no longer complex, but the price to pay is that some of the constraints (the Hamiltonian constraint) have a much more complicated
expression than the complex ones.  The parameter $\gamma$ is not relevant in the classical theory but it leads to an ambiguity
in the quantum theory that one can  compare to the $\theta$-ambiguity of QCD. 
\end{enumerate}

Contrary to Yang-Mills theory, gravity is not a gauge theory, it
is a pure constraint system and admits as symmetry group the ``huge" group of space-time diffeomorphisms supplemented with the
$SU(2)$ gauge symmetries briefly described above. The symmetries are generated in the Hamiltonian sense by the constraints: the Gauss constraints
${\cal G}_a(x)$, the vectorial constraints ${\cal H}_i(x)$ and the famous Hamiltonian or scalar constraint 
${\cal H}(x)$ where $a\in \{1,2,3\}$ are for internal or gauge indexes, $i\in \{1,2,3\}$ are for space indexes and
$x$ denotes a space point.  
Of particular interest for what follows is the symmetry group ${\cal S}={\cal G}\ltimes Diff(\Sigma)$ where ${\cal G}$
denotes the group of gauge transformations and $Diff(\Sigma)$ is the diffeomorphims group on the hyperplane $\Sigma$.
In principle, the physical phase space is obtained by first solving the constraints and second
gauge-fixing the symmetries. 

Currently, nobody knows how to construct the classical physical phase space, at least in four
dimensions, and therefore it is nonsense to hope to quantize gravity after implementation of the constraints. In Loop
Quantum Gravity, we proceed the other way around, namely quantizing the non-physical phase space before imposing the constraints.
At this point, one could ask the question why solving the quantum constraints would be simpler than solving the classical ones.
So far, we do not know any solution\footnote{In fact, we know only one solution of all the constraints when there is a cosmological
constant in the theory, known as the Kodama state \cite{kodama}. This solution was discussed several years ago \cite{kodama-discussions}
but its physical interest remains minimal.} of all the constraints even at the quantum level 
and hence no one knows precisely the physical degrees of freedom of Quantum Gravity. However, Loop Quantum Gravity provides very fascinating
intermediate results that may give a glimpse of space-time at the Planck scale \cite{discrete}, a resolution of the initial singularity for the Big Bang model \cite{bojowald}
and also a microscopic explanation of Black-Holes thermodynamics \cite{black-holes}. The problem of solving the Hamiltonian constraint is still open, but different
strategies have been developed to attack it. Recently, new results \cite{newmodel} have opened a very promising way towards its resolution...

     \subsection{The route to the quantization of gravity}
This Section is devoted to presenting the global strategy of Loop Quantum Gravity.
We have adopted the point of view of \cite{inside} which seems to us very illuminating: we start with a general discussion on the quantization
of constrained systems before discussing the case of Loop Quantum Gravity.

Starting from a symplectic (or a Poisson) manifold --- the phase space $\cal P$ ---
physicists know how to construct the associated quantum algebra. The basic idea is to promote
the classical variables into quantum operators whose non-commutative product is constructed from the classical Poisson bracket.
In that way, one constructs a quantum algebra $\mathfrak A$ whose elements are identified with (smooth) functions
on the classical phase space $\cal P$. 
The kinematical Hilbert space $\cal H$ is the carrier space of an irreducible unitary representation
of the algebra $\mathfrak A$. In the case of the quantization of a massive point particle evolving in a given potential, 
$\mathfrak A$
is the Heisenberg algebra; the kinematical Hilbert space is unique due to the famous Stone-Von Neumann theorem and the quantum states of the theory
are very well understood if the dynamics are not too complicated.
In general, $\mathfrak A$ does not admit an unique unitary irreducible representation and one has to require
some extra properties in order for $\cal H$ to be unique. For instance, it is natural to ask that symmetries are unitarily represented on $\cal H$. 
Finally, the physical Hilbert space is obtained, directly or indirectly, from solving the constraints on the kinematical Hilbert space.

Loop Quantum Gravity is based on this program. One starts with the classical phase space
${\cal P}$ which is the tangent bundle $T^*({\cal C})$ where ${\cal C}$ is the space of $SU(2)$ connections on
the  hypersurface $\Sigma$. A good ``coordinate system" for ${\cal P}$ is provided by the generators of the holonomy-flux algebra
associated to edges $e$ and surfaces $S$ of $\Sigma$ as follows:
\begin{eqnarray}
A(e) \equiv P\exp (\int_e A) \;\;\;\; \textrm{and} \;\;\;\; E_f(S) \equiv \int_S \textrm{Tr}(f\star E) \;.
\end{eqnarray}
where $f$ is a Lie algebra valued function on $\Sigma$, $\star$ is the Hodge star, $Tr$ holds for the $SU(2)$ Killing form and 
$P\exp$ is the notation for the path-ordered exponential. The symmetry group ${\cal S}={\cal G}\ltimes Diff(\Sigma)$ acts as an automorphism
of the algebra of functions on the classical algebra. The quantization of the classical algebra is straightforward and leads to the quantum
holonomy-flux algebra $\mathfrak A$. Many techniques have been used to study the representation theory of $\mathfrak A$ and the 
Gelfand'-Naimark-Segal construction is one of the most precise \cite{inside}. It consists in finding a positive state $\omega \in \mathfrak A^*$ which is 
central in the construction of the representation. Many such states exist but the requirement that $\omega$ is invariant under the action 
of $\cal S$ makes the state unique \cite{lost}. Therefore, there is an unique representation $\pi$ of the quantum holonomy-flux algebra $\mathfrak A$
which is invariant under the action of $\cal S$. This representation is the starting point of the construction of the physical states.

\subsection{Spin-networks are states of quantum geometry}
To make the representation $\pi$ of $\mathfrak A$ more concrete, let us describe its carrier space in terms 
of cylindrical functions. A cylindrical function $\Psi_{\Gamma,f}$ is defined from
a graph $\Gamma \subset \Sigma$ with $E$ edges and $V$ vertices and a function $f \in C(SU(2))^{\otimes E}$.
It is a complex valued function of the set of the holonomies $A=\{A(e_1),\cdots,A(e_E)\}$ explicitly given by:
\begin{eqnarray}
\Psi_{\Gamma,f}(A) \; = \; f(A(e_1),\cdots,A(e_E)) \;.
\end{eqnarray}
The set of cylindrical functions associated to the graph $\Gamma$ is denoted $Cyl_\Gamma$. The carrier space of the representation
$\pi$ is given by the direct and non-countable sum $Cyl(\Sigma)\equiv\oplus_\Gamma Cyl_\Gamma$ over all graphs on $\Sigma$. 
Such a sum is mathematically well-defined using the notion of a projective limit \cite{projective}. The vector space $Cyl(\Sigma)$ is endowed with a
Hilbert space structure defined from the Ashtekar-Lewandowski measure:
\begin{eqnarray}
\langle \Psi_{\Gamma,f} \vert \Psi_{\Gamma',f'} \rangle \; = \; \delta_{\Gamma,\Gamma'} \int (\prod_e d\mu(A(e))) \overline{f(A)} f'(A) 
\end{eqnarray}
where $d\mu$ denotes the $SU(2)$ Haar measure. The delta symbol means that the scalar product between two states vanishes unless they are
associated to exactly the same graph $\Gamma=\Gamma'$. This property makes the representation not weakly  continuous.
The completion of $Cyl(\Sigma)$ with respect to the Ashtekar-Lewandowski measure defines the kinematical Hilbert space $\cal H$.
It remains necessary to impose the constraints in order to extract the physical states of quantum gravity from $\cal H$. 

The Gauss constraint is quite easy to impose: a state $\Psi_{\Gamma,f}$ is invariant under the action of $\cal G$    
if the function $f$ is unchanged by the action of the gauge group on the vertices of the graph. An immediate consequence is that
the graph $\Gamma$ has to be closed. The space of gauge invariant functions is denoted ${\cal H}_0$ and is endowed with an orthonormal basis:
the basis of (gauge-invariant) spin-network states. A spin-network state $\vert S\rangle\equiv \vert \Gamma,j_e,\iota_v\rangle$ is associated to a graph $\Gamma$
whose edges $e$ are colored with $SU(2)$ unitary irreducible representations $j_e$ and vertices $v$ with intertwiners $\iota_v$ between representations
of the edges meeting at $v$. Intertwiners are generalized Clebsh-Gordan coefficients.  An example of a spin-network is given in Fig. \ref{spinnet} below.
\begin{figure}[h]
\psfrag{j}{$\ell_i$ are oriented link}
\psfrag{n}{$n_i$ are nodes}
\psfrag{j1}{$\ell_1$}
\psfrag{j2}{$\ell_2$}
\psfrag{j3}{$\ell_3$}
\psfrag{n1}{$n_1$}
\psfrag{n2}{$n_2$}
\psfrag{x}{}
\centering
\includegraphics[scale=0.6]{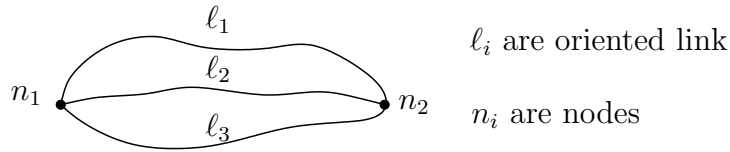}
\caption{The links are colored by representations of $SU(2)$ and the vertices by Clebsh-Gordan intertwiners.}
\label{spinnet}
\end{figure}

Imposing the diffeomorphisms constraint is also relatively easy. Roughly, it consists in identifying states whose graphs are related by a diffeomorphism
and which have the same colors once the graphs have been identified. The set of such conjugacy classes form the space ${\cal H}_{diff}$ which is endowed
with a natural Hilbert structure inherited from the Ashtekar-Lewandowski measure. Elements of ${\cal H}_{diff}$ are labelled by knots instead of graphs.

\begin{figure}[h]
\psfrag{S}{$\cal S$}
\psfrag{G}{$\Gamma$}
\psfrag{x}{$a(S)\vert S \rangle \; = \; \frac{8\pi \gamma \hbar G}{c^3} \sum_{e \; \textrm{crosses}\; S} \sqrt{j_e(j_e+1)} \vert S\rangle $}
\psfrag{t}{}
\centering
\includegraphics[scale=0.7]{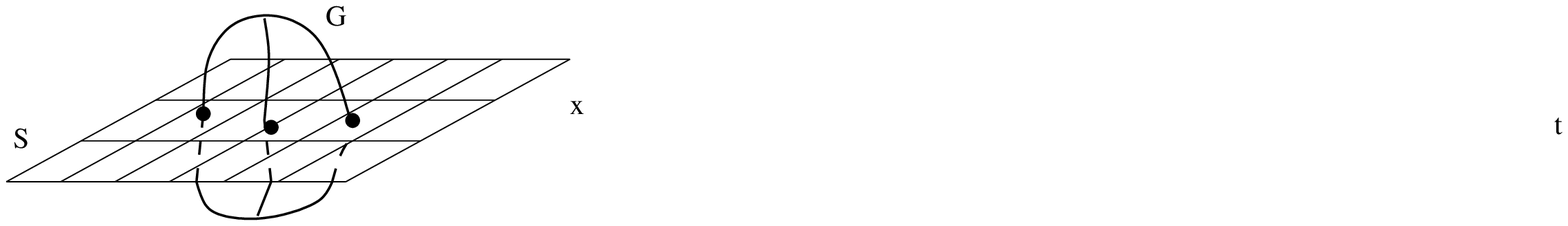}
\caption{Illustration of the action of the area operator on a given spin-network.}
\label{areaspectrum}
\end{figure}

Before discussing the remaining constraint in the next Section, let us give the physical interpretation of a spin-network state.
To do so, we need to introduce some geometrical operators, such as those that relate to the area $a(S)$ of a surface $S$ and the volume $v(R)$ of a domain $R$.
The classical expressions of $a(S)$ and $v(R)$ are functions on the $E$-field given by \cite{discrete}:
\begin{eqnarray}
a(S) = \int_S \!\! d^2x\sqrt{E_i^a E_j^b n_an_b}\;\;\;
\textrm{and} \;\;\;
v(R)=\int_R\!\!d^3x \sqrt{\frac{\vert \epsilon_{abc}\epsilon_{ijk} E^{ai}E^{bj}E^{ck}\vert}{3!}}
\end{eqnarray}
where $n_a$ denotes the normal of the surface $S$ and $\epsilon_{abc}$ are the totally antisymmetric tensors.
To promote these classical functions into quantum operators acting on the kinematical states, one has to introduce
 regularizations for the area or the volume due to the presence of the square roots in the previous classical definitions.
There exist therefore some ambiguities in the definition of the quantum geometrical operators, above all in the case of
the volume.  For the area, the standard regularization leads to an operator 
$a(S)$ whose action on a spin-network state $\vert S \rangle$ is illustrated in Fig. \ref{areaspectrum},
where the sum runs over all the edges of the graph $\Gamma$ associated to $\vert S \rangle$ that cross the surface $S$.
We have assumed that the edges always cross $S$ transversely; the formula can be generalized for other, more general, cases \cite{discrete}. We have explicitly
introduced all the fundamental constants in order to show, in particular, the dependence on the Immirzi parameter $\gamma$ \cite{immirzi}. We also see
immediately that spin-network states are eigenstates of $a(S)$ with discrete eigenvalues. 
A similar but much more involved result exists for the volume operator $v(R)$: it acts on the nodes
of the spin-network states and also has a discrete spectrum.
As a result, at the kinematical level, space appears discrete in Loop Quantum Gravity.

\subsection{The problem of the Hamiltonian constraint}\label{dynamics}
Solving the Hamiltonian constraint is still an open issue. Two main roads have been developed to understand this constraint: 
the master program \cite{master} and Spin-Foam models \cite{spinfoam}. The master program, initiated and mainly developed by Thiemann, is an attempt to
regularize the Hamiltonian constraint in order to find its kernel. Even if we still do not have a precise description of the
physical Hilbert space, Thiemann proved an existence theorem that ensures physical states exist.  We will not discuss this
approach further here.

Spin-Foam models are an alternative attempt to solve the dynamics from a covariant point of view. The idea consists in
finding the physical scalar product between spin-network states not necessarily solutions of the Hamiltonian constraint.
Of course, the two problems are closely related. The physical scalar product should be given  by the path integral  of gravity,
if one could give a meaning to this. Spin-Foam models are precisely proposals for the path integral of gravity.
These proposals are based on the Plebanski formulation of gravity where gravity is described as a constrained $BF$
theory. All $BF$ theories are topological theories whose path integral can be easily (and formally) written in terms of
combinatorial objects that we do not want to describe here.  One starts with this path integral and tries to impose the constraints
that make gravity a $BF$ theory at the level of the path integral. For the moment, there is no precise implementation of 
the constraints, but there do exist proposals.
Recently a promising new Spin-Foam model has been described \cite{newmodel}. 

\begin{figure}[h]
\psfrag{a}{${\cal A}=\langle S \vert S' \rangle_{phys}$}
\centering
\includegraphics[scale=0.7]{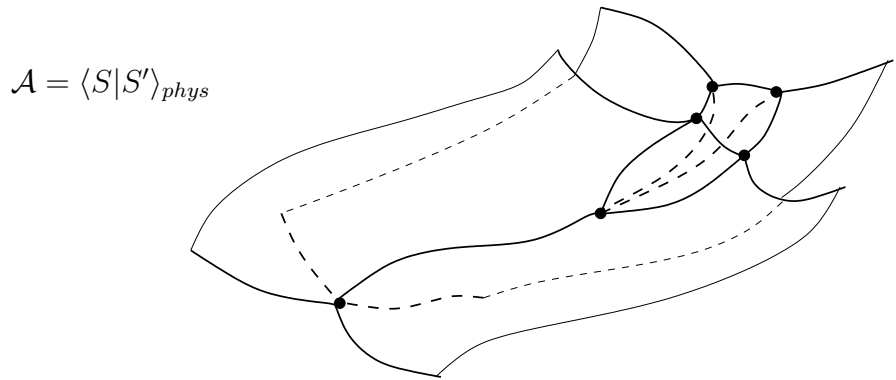}
\caption{Spin-Foam models propose an amplitude to each graph interpolating two given states. This amplitude
is related to the physical scalar product between the two states.}
\label{spinfoam}
\end{figure}

In the context of Spin-Foam models, the physical scalar product between two spin-network states is given by a certain 
evaluation of topological graphs interpolating the two graphs defining the two spin-networks, as illustrated
in Fig. \ref{spinfoam}.  
The rules for computing the 
amplitude of the graph are model dependent and can be viewed as generalizations of the Feynman rules for standard QFT.
We may be far from having a clear and complete description of the physical Hilbert space of quantum gravity but the road proposed in LQG is very fascinating.

\section{Three dimensional Euclidean quantum gravity}
Let us underline two aspects, among the most important, concerning Loop Quantum Gravity in four dimensions.
The first one is the possibility that space is discrete at the Planck scale. The second one is the difficulty
in solving the dynamics of quantum gravity, and the subsequent impossibility of identifying the physical states. 
Thus, to test the discreteness of space, one could use a simple toy model where the dynamics is easy to solve, and which already exhibits 
a discreteness of space. Three dimensional Euclidean quantum gravity offers an ideal framework in this regard.

\subsection{Construction of the non-commutative space}
Anyone who would claim to have quantized gravity should, at the very least, be able to give a precise meaning to the 
formal expression for the path integral
\begin{eqnarray}\label{pathintegral}
{\cal Z} \; = \; \int [{\cal D}g][{\cal D}\varphi] e^{iS[g,\varphi]}
\end{eqnarray}
where $S[g,\varphi]$ is the Einstein-Hilbert action for the metric $g$ coupled to any matter field $\varphi$.
The difficulty in performing such an integral is obviously hidden in the construction of a suitable measure $[{\cal D}g]$ for
the space of metrics modulo diffeomorphisms. 

If one uses standard perturbative techniques to compute (\ref{pathintegral}), namely, one first writes $g=\eta + h$ 
as the sum of the flat metric $\eta$ and a ``fluctuation" $h$, then performs the integration over the variable 
$\varphi$ on the flat metric and finally sums over all the fluctuations $h$, one gets into trouble because the theory
is non-renormalizable. Furthermore, this method strongly breaks the covariance of the theory by specifying one 
background metric, and so appears not to be well adapted to general relativity.

As was mentioned in the introduction, in order to circumvent these difficulties, 
one could try the other way around, performing first the integration over the gravitational degrees of freedom.
This idea makes sense for 3D Euclidean gravity, which can be completely quantized by different techniques. 

\subsubsection{Quantum Gravity and non-commutativity}
Of particular interest is the Spin-Foam framework which gives tools for performing, at least formally, the integration
over the metric variable in (\ref{pathintegral}). Indeed, it has been argued that the path integral (\ref{pathintegral}) reduces to a path
integral of an effective quantum field theory $S_{eff}[\varphi]$ as follows \cite{fl}
\begin{eqnarray}\label{effectivefield}
{\cal Z} \; = \; \int [{\cal D}\varphi] e^{iS_{eff}[\varphi]}
\end{eqnarray}
where $S_{eff}[\varphi]$ defines the action of a sole field $\varphi$ on a fixed, but non-commutative, background. 
The non-commutative space is a deformation of the classical flat Euclidean space whose deformation parameter is the Planck
length $\ell_P=\sqrt{G\hbar/c^3}$. 
Thus, quantum gravity would make ``space-time" non-commutative, at least when space-time is three dimensional and Euclidean.
Now, we aim at giving a precise definition of this non-commutative space. Before going into details of the definition,
let us emphasize that this non-commutative space is unrelated to the particular Moyal non-commutative space \cite{moyal} that appears within the String Theory framework.

The path integral approach to constructing the non-commutative geometry we have just outlined is certainly the most appealing
at a conceptual level. Nonetheless, we will adopt here a more ``canonical" way which is, at a technical level, simpler and also
quite intuitive \cite{moi1, jihad}. Our starting point is the fact that the classical symmetry group of the theory is deformable into a quantum group.
It is indeed well-known that quantum groups play a crucial role in the quantization of three dimensional gravity; the link between 
quantum gravity and knot invariants in three dimensional manifolds \cite{witten1} is certainly one of the most beautifull illustrations of this fact.

Three dimensional gravity, for all values of the cosmological constant $\Lambda$ and whatever the signature of space-time, 
is an exactly solvable system, as pointed out by Witten \cite{witten2}. It can be reformulated as a Chern-Simons theory, which is a gauge theory
whose gauge group is of the form $C^\infty(M,G)$, $M$ being the space-time and $G$ a Lie group. For $\Lambda=0$ and Euclidean signature,
the group $G=ISU(2)\equiv SU(2) \ltimes \mathbb R^3$ is the (universal cover of the) isometry group of the three dimensional flat Euclidean 
space. This group gets deformed when the theory is quantized \cite{muller}. Only an idea of the deformation is given in the following Section where we hope 
the reader gets at least the physical content of the deformation process. Mathematical and technical details can be found in \cite{jihad} for instance. 

\subsubsection{The Quantum Double plays the role of the isometry algebra}
In the combinatorial quantization scheme \cite{combinatorial}, the deformation of the isometry group is very clear. Classical groups are turned into quantum groups
and the construction of the quantum physical states uses as a central tool the representation theory of these quantum groups. In the case we are
interested in, the quantum group is the Drinfeld double of $SU(2)$, called also the quantum double or the double for short
and denoted $DSU(2)$. The notion of quantum double
is very general in the sense that it is possible to construct the quantum double $DA$ for any Hopf algebra $A$. $DSU(2)$ is in fact the quantum double
of the commutative algebra $C(SU(2))$ of smooth functions on $SU(2)$ which is endowed with a Hopf algebra structure:
the algebra is defined by the point-wise product of functions and the co-algebra is determined by the standard co-product $\Delta: C(SU(2))\rightarrow
C(SU(2))\otimes C(SU(2))$ such that $\Delta(f)(a,b)=f(ab)$ for any group elements $a,b\in SU(2)$. 
The detailed definition of $DSU(2)$ can be found in several references \cite{moi1, jihad} but we only need to mention that, as a vector space, $DSU(2)$ is the tensor product $C(SU(2))\otimes C[SU(2)]$
where $C[G]$ denotes the group algebra of $G$, i.e. the algebra of formal linear combination of elements of $G$. In particular, $G$ is a subset of
$\mathbb C[G]$.

The double $DSU(2)$ is, precisely, a deformation of the algebra $\mathbb C[ISU(2)]$. In fact, the deformation concerns only the co-algebra 
structure which is a central notion in constructing tensor products of representations. There exists an algebra morphism between $DSU(2)$ and 
$\mathbb C[ISU(2)]=C[\mathbb R^3]\otimes C[SU(2)]$, more precisely $DSU(2)$ is included as an algebra into $\mathbb C[ISU(2)]$. 
The $\mathbb C[SU(2)]$ part of $DSU(2)$ is identified
to $\mathbb C[SU(2)]\subset \mathbb C[ISU(2)]$ whereas $\mathbb C[\mathbb R^3]$ is sent to $C(SU(2))$. Thus, if one trivially identifies
$\mathbb C[\mathbb R^3]$ with the algebra of functions $C(\mathbb R^3)$ on the Euclidean space $\mathbb R^3$, then the deformation process transforms
$C(\mathbb R^3)$ into $C(SU(2))$. Roughly, the deformation works as a compactification of the space $\mathbb R^3$ which becomes the space $SU(2)$ that
can be identified to the sphere $S^3$. One understands that such a compactification needs a parameter with dimension of a length and here the
Planck length $\ell_P$ enters. In other words, the Planck length is crucial for transforming momentum vectors $\vec v$ in $\mathbb R^3$ into group elements $\nu(\vec{v})$
according to the formula $\nu(\vec v)=e^{i\ell_P \vec{v}\cdot \vec{\sigma}} \in SU(2)$, where the notation $\vec{\sigma}=(\sigma_1,\sigma_2,\sigma_3)$ 
holds for the generators of the Lie algebra $\mathfrak{su}(2)$. 

In brief, the quantum double $DSU(2)$ is a deformation of the group algebra $\mathbb C[ISU(2)]$
where the rotational part is not affected by the deformation and the translational part is 
compactified in the sense described above. The group of translations becomes compact and also 
non-commutative. This is the fundamental reason why space-time becomes non-commutative.

\subsubsection{The Quantum Geometry defined by its momenta space}
The quantum geometry at the Planck scale is defined as the space which admits the quantum double as an 
isometry algebra. This definition is analogous to the classical one: indeed, the classical flat  Euclidean
space $\mathbb E^3$ admits the Euclidean group $ISU(2)$ as an isometry group and moreover can be described as
the quotient $ISU(2)/SU(2)$. At the non-commutative level, one has to adapt such a construction (by quotient), for
the non-commutative space is defined indirectly by its algebra of functions $\cal A$. However, the construction is quite
easy to generalize and leads to the fact that $\cal A$ is the convolution algebra of $SU(2)$ distributions \cite{jihad}, denoted
\begin{eqnarray}\label{defofA}
{\cal A} \; \equiv \; (C(SU(2))^*,\circ) \;.
\end{eqnarray} 
 This algebra is trivially non-commutative.
The algebra of functions $C(SU(2))$ endowed with the convolution product is a particular sub-algebra of $\cal A$
and  the product of two functions is explicitely given by:
\begin{eqnarray}
(f_1\circ f_2)(a)=\int d\mu(x) f_1(x)f_2(x^{-1}a),
\end{eqnarray}
where $d\mu(x)$ is the $SU(2)$ Haar measure.

The algebra $\cal A$ admits different equivalent formulations which have distinct physical interpretations.
The formulation in (\ref{defofA}) above is called the momentum representation: it is indeed a deformation of the commutative algebra $C(\mathbb R^3)$
of distributions  on the tangent space $\mathbb R^3$ of $\mathbb E^3$. At the Planck scale, the momenta become
grouplike. 

By construction, $\cal A$ provides a representation space of $DSU(2)$ which can be interpreted, in this way,
as a symmetry algebra of $\cal A$ whose action will be denoted $\rhd$.
More precisely, translation
elements are functions on $SU(2)$ and act by multiplication on $\cal A$ whereas rotational elements are 
$SU(2)$ elements and act by the adjoint action:
\begin{eqnarray}
\forall \; \phi \in {\cal A} \;\;\; f \rhd \phi = f \phi \;\;\; \textrm{and} \;\;\; 
u \rhd \phi = \textrm{Ad}_u \phi \;. 
\end{eqnarray}
The adjoint action is defined by the relation $\langle f,\textrm{Ad}_u\phi \rangle=
\langle \textrm{Ad}_{u^{-1}}f,\phi \rangle$ with $\textrm{Ad}_u f(x)$ given by $f(u^{-1}xu)$ for any $u,x$ in $SU(2)$.

\subsubsection{The fuzzy space formulation}
Thus, we have a clear definition of the deformed space of momenta.
To get the quantum analogue of the space $C(\mathbb E^3)$ itself, 
we need to introduce a Fourier transform on $C(SU(2))^*$. This is done by making use of harmonic analysis 
on the group $SU(2)$:
the Fourier transform of a given $SU(2)$-distribution is the decomposition of that distribution into 
(the whole set or a subset of) 
unitary irreducible representations (UIR) of $SU(2)$. These UIR are labelled by a spin $j$, they are finite dimensional
of dimension $d_j=2j+1$.  The Fourier transform is an algebra morphism which is explicitly defined by:
\begin{eqnarray}
{\cal F}: \;\; C(SU(2))^* & \longrightarrow & \textrm{Mat}(\mathbb C) \equiv 
\bigoplus_{j=0}^\infty \textrm{Mat}_{d_j}(\mathbb C)  \\
\phi & \longmapsto & \widehat{\Phi}\equiv{\cal F}[\phi]= \oplus_j {\cal F}[\phi]^j \, = \, \oplus_j (\phi \circ D^j)(e)
\end{eqnarray}
where $\textrm{Mat}_{d}(\mathbb C)$ is the set of $d$ dimensional complex matrices, $D^j_{mn}$ are
the Wiegner functions and $\circ$ is the convolution product.
When $\phi$ is a function, its Fourier matrix components are obtained by performing the following integral 
\begin{eqnarray}
{\cal F}[\phi]^j_{mn} \; \equiv \; \int d\mu(u) \, \phi(u) \, D^j_{mn}(u^{-1}) \,.
\end{eqnarray}
The inverse map ${\cal F}^{-1}:\textrm{Mat}(\mathbb C)\rightarrow C(SU(2))^*$ associates to any family of matrices
$\widehat{\Phi}=\oplus_j \widehat{\Phi}^j$ a distribution according to the formula:
\begin{eqnarray}
\langle f, {\cal F}^{-1}[\widehat{\Phi}] \rangle \; = \;  \sum_j d_j 
\int d\mu(u) \overline{f(u)} \, \textrm{tr}(\widehat{\Phi}^j\;D^j(u)) \, 
\equiv \, \int d\mu(u) \overline{f(u)} \,\textrm{Tr}(\widehat{\Phi}D(u)){\rm~~~~~~~} 
\end{eqnarray}
for any function $f \in C(SU(2))$. We have introduced the notations $D=\oplus_j D^j$ and 
$\textrm{Tr}\widehat{\Phi}=\sum_j d_j \textrm{tr}(\widehat{\Phi}^j)$.
Therefore, it is natural to interpret the algebra 
$\textrm{Mat}(\mathbb C)$ as a deformation of the classical algebra 
$C(\mathbb E^3)$ and then 
three dimensional Euclidean quantum geometry is fundamentally non-commutative and fuzzy.

\subsubsection{Relation to the classical geometry}
It is not completely trivial to show how the algebra of matrices $\textrm{Mat}(\mathbb C)$ is a deformation of
the classical algebra of functions on $\mathbb E^3$. 

To make it more concrete, it is necessary to construct a precise link between $C(SU(2))^*$ and
$C(\mathbb R^3)^*$ for the former space is supposed to be a deformation of the latter. First,
we remark that it is not possible to find a vector space isomorphism between them because $SU(2)$
and $\mathbb R^3$ are not homeomorphic: in more physical words, there is no way to establish a 
one-to-one mapping between distributions on $SU(2)$ and distributions on $\mathbb R^3$, for $SU(2)$ and
$\mathbb R^3$ have different topologies. Making an explicit link between these two spaces is in fact
quite involved and one construction has been proposed in  \cite{jihad}. The aim of this 
Section is to recall only the main lines of that construction; more details can be found in 
\cite{jihad}. For pedagogical reasons, we also restrict the space $C(SU(2))^*$ to its subspace  
$C(SU(2))$ and then we are going to present the link between $C(SU(2))$ and $C(\mathbb R^3)$.
\begin{enumerate}
\item First, we need to introduce a parametrization of $SU(2)$: $SU(2)$ is identified with 
$S^3=\{(\vec{y},y_4)\in \mathbb R^4 \vert y^2 + y_4^2=1\}$ and any $u\in SU(2)$ is given by
\begin{eqnarray}
u(\vec{y},y_4)\, = \, y_4-i\vec{y}\cdot \vec{\sigma}
\end{eqnarray}
in the fundamental representation in terms of the Pauli
matrices $\sigma_i$. For later convenience, we cut $SU(2)$ in two parts: the northern hemisphere $U_+$ ($y_4>0$) 
and the southern hemisphere $U_-$ ($y_4<0$).
\item Then, we construct bijections between the spaces $U_\pm$ and the open ball of $\mathbb R^3$
$B_{\ell_P}=\{\vec{p} \in \mathbb R^3 \vert p<\ell_P^{-1}\}$: to each element $u\in U_\pm$ we 
associate a vector $\vec{P}(u)=\ell_P^{-1}\vec{y}$. 
These bijections implicitly identify $\vec{P}(u)$ with the physical momenta of the theory.
Note that this is a matter of choice: on could have chosen another expression for $\vec P(u)$ and there are no
physical arguments to distinguish one from the other. We have made what seems to be, for various different reasons, the
most natural and convenient choice.
\item As a consequence, any function $\phi\in C(SU(2))$ is associated to a pair of functions 
$\phi_\pm \in C(U_\pm)$, themselves being associated, using the previous bijections, to a pair
of functions $\psi_\pm \in C_{B_{\ell_P}}(\mathbb R^3)$ which are functions on $\mathbb R^3$ with support
on the ball $B_{\ell_P}$. In that way, we construct two mappings 
$\mathfrak a_\pm:C(U_\pm)\rightarrow C_{B_{\ell_P}}(\mathbb R^3)$ such that $\mathfrak a_\pm(\phi_\pm)=\psi_\pm$
are explicitly given by:
\begin{eqnarray}
\psi_\pm(\vec{p})\!=\!\!\int\!\!d\mu(u) \delta^3(\vec p\!-\!\vec P(u)) \phi_\pm(u)\!=\!\frac{v_{\ell_P}}{\sqrt{1\!-\!{\ell_P^2 p^2}}}
\phi(u({\ell_P\vec{p}},\pm\sqrt{1\!-\!{\ell_P^2 p^2}})),{\rm~~~~~~~}
\end{eqnarray}
where $v_{\ell_P}=\ell_P^3/2\pi^2$.
We have thus established a vector space isomorphism $\mathfrak a=\mathfrak a_+ \oplus \mathfrak a_-$
between $C(SU(2))$ and $C_{B_{\ell_P}}(\mathbb R^3)\oplus C_{B_{\ell_P}}(\mathbb R^3)$. We need two functions
on $\mathbb R^3$ to characterize one function of $C(SU(2))$. The mapping $\mathfrak a_\pm$ satisfies the important
following property: the action of the Poincar\'e group $ISU(2)\subset DSU(2)$ on $C_{B_{\ell_P}}(\mathbb R^3)$
induced by the mappings $a_\pm$ is the standard covariant one, namely
\begin{eqnarray}
\xi \rhd \mathfrak a_\pm(\phi_\pm) \; = \; \mathfrak a_\pm(\xi \rhd \phi_\pm) \;\;\;\;\;\;
\forall \, \xi \in ISU(2) \subset DSU(2) \;.
\end{eqnarray}
In the r.h.s. (resp. l.h.s.), $\rhd$ denotes the action of $\xi \in ISU(2)$ (resp. $\xi$ viewed as an element of 
$DSU(2)$) on $C(\mathbb R^3)$ (resp. $C(SU(2))$).
This was, in fact, the defining property of the mappings $\mathfrak a_\pm$.
\end{enumerate}
Now, we have a precise relation between $C(SU(2))$ and $C(\mathbb R^3)$. Using the standard Fourier transform
$\mathfrak F:C(\mathbb R^3)^*\rightarrow C(\mathbb E^3)$ restricted to $C_{B_{\ell_P}}(\mathbb R^3)$, one obtains
the following mapping:
\begin{eqnarray}
\mathfrak m \, \equiv \, \mathfrak F \circ \mathfrak a \; : \; C(SU(2)) \, \longrightarrow \, C_{\ell_P}(\mathbb E^3)
\end{eqnarray}
where $C_{\ell_P}(\mathbb E^3)$ is defined as the image of $C(SU(2))$ by $\mathfrak m$. 
It will be convenient to introduce the obvious notation $\mathfrak m=\mathfrak m_+ \oplus \mathfrak m_-$.
We have the vector space isomorphism 
$C_{\ell_P}(\mathbb E^3) \simeq \widetilde{C}_{B_{\ell_P}}(\mathbb R^3) \oplus \widetilde{C}_{B_{\ell_P}}(\mathbb R^3)$,
where $\widetilde{C}_{B_{\ell_P}}(\mathbb R^3)$ is the subspace of functions on $\mathbb E^3$ whose spectra are strictly
contained in the open ball $B_{\ell_P}$ of radius $\ell_P^{-1}$. Elements of $C_{\ell_P}(\mathbb E^3)$ 
are denoted $\Phi_+\oplus \Phi_-$ where $\Phi_\pm(x) \in  \widetilde{C}_{B_{\ell_P}}(\mathbb R^3)$. 
The explicit relation between $C(SU(2))$ and $C_{\ell_P}(\mathbb E^3)$ is
\begin{eqnarray}
\Phi_\pm(x) \, \equiv \, \mathfrak m_\pm(\phi_\pm)(x) \, = \, \int d\mu(u) \phi_\pm(u) \, \exp(iP(u)\cdot x) \;.
\end{eqnarray}
This transform is clearly invertible.

\medskip

It remains to establish the link between $C_{\ell_P}(\mathbb E^3)$ and the space of matrices 
$\textrm{Mat}(\mathbb C)$. To do so, we make use of the mapping $\cal F$ between $C(SU(2))$ and $\textrm{Mat}(\mathbb C)$
and the mapping $\mathfrak m$ between the same $C(SU(2))$ and $C_{\ell_P}(\mathbb E^3)$. If we denote by 
$\widehat{\Phi}_\pm$ the images of $\phi_\pm$ by $\cal F$, then we have:
\begin{eqnarray}\label{relPhiMat}
\Phi_\pm(x) \; = \; \textrm{Tr}(K_\pm^\dagger(x)\widehat{\Phi}_\pm)
\end{eqnarray}
where $K_\pm$ can be interpreted as the components of the element 
$K\equiv K_+\oplus K_- \in \textrm{Mat}(\mathbb C) \otimes C_{\ell_P}(\mathbb E^3)$  defined by the integral:
\begin{eqnarray}\label{def de K}
K_\pm(x) \, \equiv \, \int_{U_\pm} d\mu(u) \, D(u) \, \exp(-iP(u)\cdot x) \;.
\end{eqnarray}
\medskip
The relation (\ref{relPhiMat}) is invertible. One can interpret the functions $\Phi_\pm(x)$ as a kind of continuation
to the whole Euclidean space of the discrete functions $\widehat{\Phi}^j_{\pm mn}$ which are a priori defined only on
a infinite but enumerable set of points. Given $x \in \mathbb E^3$, each matrix element $\widehat{\Phi}^j_{\pm mn}$
contributes to the definition of $\Phi_\pm(x)$ with a complex weight $\overline{K^j_{\pm nm}(x)}$.

\medskip

For the moment, we have only described the vector space structure of $C_{\ell_P}(\mathbb E^3)$.
However, this space inherits a non-commutative algebra structure when we ask the mapping $\mathfrak m$
to be an algebra morphism. The product between two elements $\Phi_1$ and $\Phi_2$ in $C_{\ell_P}(\mathbb E^3)$
is denoted $\Phi_1\star\Phi_2$ and is induced from the convolution product $\circ$ on $C(SU(2))$ as follows:
\begin{eqnarray}
\Phi_1\star\Phi_2 \; = \; \mathfrak m(\mathfrak m^{-1}(\Phi_1) \circ \mathfrak m^{-1}(\Phi_2))\;.
\end{eqnarray}
The $\star$-product is a deformation of the classical pointwise product. 

In order to make the $\star$-product more intuitive, 
it might be useful to consider some examples of products of functions. 
The most interesting functions to consider first are surely the plane waves.
Unfortunately, plane waves are not elements of $C(SU(2))$ but are pure distributions and hence,
their study goes beyond what we have covered in this paper. Nevertheless, we will see that it is 
possible to extend the previously presented results to the case of the plane waves with some assumptions.
Plane waves are defined as eigenstates of the generators $P_a$ and then, as we have already underlined,
a plane wave is represented by the distribution
$\delta_u$ with eigenvalue $P_a(u)$ which is interpreted as the momentum of the plane wave. 
Plane waves are clearly degenerate, as $P_a(u)$ is not invertible in $SU(2)$: this result illustrates 
the fact that we need two functions $\Phi_+\oplus \Phi_- \in C_{\ell_P}(\mathbb E^3)$ to characterize one function 
$\phi \in C(SU(2))$.
The representations
of the plane wave in the matrix space $\textrm{Mat}(\mathbb C)$ and in the continuous space
${C}_{\ell_P}(\mathbb E^3)$ are respectively given by:
\begin{eqnarray}
{\cal F}(\delta_u)^j \; = \; D^j(u){}^{-1} \;\;\;\; \textrm{and}  \;\;\;\;
{\mathfrak m}(\delta_u)(x) \; \equiv \; w_u(x)
\end{eqnarray}
where $w_u(x)=\exp(iP_a(u)x^a) \oplus 0$ if $u\in U_+$ and $w_u(x)=0 \oplus \exp(iP_a(u)x^a)$ if $u\in U_-$.
The framework we have described does not include the case $u\in \partial U_+=\partial U_-$ which is nonetheless
completely considered in \cite{jihad}.
The $\star$-product between two plane waves reads:
\begin{eqnarray}
w_u \; \star \; w_v \; = \; w_{uv}
\end{eqnarray}
if $u$, $v$ and $uv$ belong to $U_+$ or $U_-$. This product can be trivially extended
to the cases where the group elements belong to the boundary $\partial U_+=\partial U_-$.
As a result, one interprets $P_a(u) \boxplus P_a(v) \equiv P_a(uv)$ as the deformed addition rule for 
momenta in the non-commutative space.

Other interesting examples to consider are the coordinate functions. 
They are easily defined using the plane waves and their definition in the $C(SU(2))^*$ and
$\textrm{Mat}(\mathbb C)$ representations are:
\begin{eqnarray}
\chi_a =2i\ell_P \xi_a \delta_e \in C(SU(2))^* \;\;\;\;\;\;\;\;
\widehat{x}_a = 2\ell_P D(J_a) \in \textrm{Mat}(\mathbb C)
\end{eqnarray}
where $\xi_a$ is the $SU(2)$ left-invariant vector field and $J_a$ the generators of the $\mathfrak{su}(2)$
Lie algebra satisfying $[J_a,J_b]=2i\epsilon_{ab}{}^c J_c$. In the $C_{\ell_P}(\mathbb E^3)$ representation,
the coordinates are given by $X_a\equiv(x_a\oplus 0)$; only the first component is non-trivial.
It becomes straightforward to show that the coordinates satisfy the relation
\begin{eqnarray}
[X_a,X_b]_\star \; \equiv \; X_a  \star X_b  - 
X_b \star X_a \; =\; i\ell_P \epsilon_{ab}{}^c X_c 
\end{eqnarray}
and therefore do not commute, as expected.

\subsection{Constructing the quantum dynamics}
In this Section, we introduce some mathematical tools for defining the dynamics in the non-commutative space ---
an integral in order to define an action, and a derivative operator in order to define the kinematical energy of
the system.

\subsubsection{An integral on the quantum space to define the action}
An important property is that the non-commutative space admits an invariant measure 
$h:C \rightarrow \mathbb C$. To be more precise, $h$ is well defined  
on the restriction of $C \simeq C(SU(2))^*$ to $C(SU(2))$.
The invariance is defined with
respect to  the symmetry action of the Hopf algebra $DSU(2)$. Let us give the expression of
this invariant measure in the different formulations of the non-commutative space:
\begin{eqnarray}\label{measure}
h(\phi) \; = \; \phi(e) \; = \; \textrm{Tr}(\widehat{\Phi}) \; = \; 
 \int \frac{d^3x}{(2\pi)^3v_{\ell_P}} \; \Phi_+(x) 
\end{eqnarray}
where $\phi \in C(SU(2))$, $\widehat{\Phi}={\cal F}[\phi]$ and $\Phi_+(x)={\mathfrak m}_+[\phi](x)$.
Note that $\int d^3x$ is the standard Lebesgue measure on the classical manifold $\mathbb E^3$. 
Sometimes, such a measure is called a trace. It permits us to define a norm on the algebra $C$ from
the hermitian bilinear form 
\begin{eqnarray}
\langle \phi_1,\phi_2\rangle \; \equiv \; h({\phi}^{\flat}_1\phi_2) \; = \; 
\int d\mu(u) \, \overline{\phi_1(u)}\phi_2(u)
\end{eqnarray}
where $\phi^{\flat}(u)=\overline{\phi(u^{-1})}$. 

\subsubsection{Derivative operators to define the dynamics}
Derivative operators $\partial_\xi$ can be deduced from the action of infinitesimal translations: 
given a vector $\xi \in \mathbb E^3$, we have $\partial_\xi=\xi^a\partial_a$
where $\partial_a=iP_a$ is the translation operator we have introduced in the previous section.
When acting on the $C(SU(2))$ representation, $\partial_\xi$ is the multiplication
by the function $i\xi^aP_a$; it is the standard derivative
when acting on the continuous ${C}_{\ell_P}(\mathbb E^3)$ representation (using the mapping $\mathfrak m$); 
finally it is a finite difference operator
when acting on the fuzzy space representation $\textrm{Mat}(\mathbb C)$ (using the Fourier transform $\cal F$).
Its expression in the matrix representation is then given by:
\begin{eqnarray}\label{derivative}
(\partial_a \widehat{\Phi})^j_{st} \; = \; -\frac{1}{\ell_P d_j} D^{1/2}_{pq}(J_a) && 
\!\!\!\!\!( \sqrt{(j+1+2qs)(j+1+2tp)} \; \widehat{\Phi}^{j+1/2}_{q+s \, p+t} \nonumber\\
&& + (-1)^{q-p}\sqrt{(j-2qs)(j-2pt)} \; \widehat{\Phi}^{j-1/2}_{q+s \, p+t} ).
\end{eqnarray} 
The interpretation of the formula (\ref{derivative}) is clear. Note however an important point: 
the formula (\ref{derivative}) defines a second order operator in the sense that it involves
$\widehat{\Phi}^{j-1/2}$ and $\widehat{\Phi}^{j+1/2}$ that are not nearest matrices but second nearest matrices.

The derivative operator is obviously necessary for defining a dynamics in the non-commutative fuzzy space.
The ambiguity in the definition of $P_a$ implies immediately an ambiguity in the dynamics. 
For instance, the fact that $C_a(j,k)$
relates matrices $\widehat{\Phi}^j$ with $\widehat{\Phi}^{j\pm 1/2}$ only is a consequence
 of the choice of $P_a$ which is in fact a function
whose non-vanishing Fourier modes are the matrix elements of a dimension 2 matrix: 
indeed, $P_a(u)=\ell_P^{-1}\textrm{tr}_{1/2}(J_au)$. 
Another choice would lead to a different dynamics and then there is an ambiguity. 
Such ambiguities exists as well in full Loop Quantum Gravity \cite{ambiguity}.

\subsubsection{Free field: solutions and properties}
Now, we have all the ingredients to study dynamics on the quantum space.
Due to the fuzzyness of space, equations of motion will be discrete and therefore, there is in general no equivalence
between the Lagrangian and Hamiltonian dynamics. Here, we choose to work in the Euler-Lagrange point of view, i.e.
the dynamics is governed by an action of the type:
\begin{eqnarray}\label{action}
S_\star[\Phi,J] \; = \; \frac{1}{2}
\int \frac{d^3x}{(2\pi)^3 v_{\ell_P}} \; 
\left(\partial_\mu \Phi \star \partial_\mu \Phi \; + \; V(\Phi,J)\right)_+(x)
\end{eqnarray}
where $V$ is the potential that depends on the field $\Phi$ and eventually on some exterior fields $J$.
The action has been written in the ${C}_{\ell_P}(\mathbb E^3)$ formulation to mimic easily the classical situation.

Obviously, finding the equations of motions reduces to extremizing the previous action, but with the constraint 
that $\Phi$ belongs to ${C}_{\ell_P}(\mathbb E^3)$: in particular, $\Phi$ (as well as the exterior field) 
admits two independent components 
$\Phi_\pm$ which are classical functions on $\mathbb E^3$ whose spectra are bounded. The action 
(\ref{action}) couples these two components generically.
Even when one of the two fields vanishes, for instance $\Phi_-=0$, 
it happens in general that the extrema of the functional
$S[\Phi]$ differ from the ones that we obtain for a classical field $\Phi$ whose action would be formally the same
functional but defined with the pointwise product instead of the $\star$ product. This makes the classical solutions
in the deformed and non-deformed cases different in general.
Let us state this point more precisely. When the field is free, in the sense that $V$ is quadratic (with a mass term), deformed solutions
are the same as classical ones. However, solutions are very different when the dynamics
is non-linear, and the differences are physically important. 

\medskip

First, let us consider the case of a free field: we assume that $V(\Phi)=\mu^2\Phi\star \Phi$
where $\mu$ is a positive parameter. 
Equations of motion are:
\begin{eqnarray}
{\Delta \widehat{\Phi}}^j\; + \; \mu^2 \widehat{\Phi}^j \; = \; 0 \;\;\; \textrm{for all spin $j$}.
\end{eqnarray}
Due to the quite complicated expression of the derivative operator, it appears more convenient
to solve this set of equations in the $C(SU(2))$ representation. 
Indeed, these equations are equivalent to the fact that $\phi ={\cal F}^{-1}[\Phi]$ 
has support in the conjugacy classes $\theta \in [0,2\pi[$
such that $\sin^2(\theta/2)=\ell_P^2\mu^2$. Thus, a solution exists only if $\mu \leq \ell_P^{-1}$, 
in which case we write $\mu=\ell_P^{-1}\sin (m/2)$  with $0<m<\pi$.
Then the solutions of the previous system are given by $\widehat{\Phi}=\widehat{\Phi}_++\widehat{\Phi}_-$ with: 
\begin{eqnarray}
\widehat{\Phi}_\pm^j \; = \; \int d\mu(u) \; \mathbb I^\pm_m(u) \left( \alpha(u) D^j(u) \; + \; 
\beta(u) D^j(u)^\dagger \right) 
\end{eqnarray}
where $\alpha$ and $\beta$ are $SU(2)$ complex valued functions;
the notation $\mathbb I^\pm_m$ holds for the characteristic functions on the conjugacy class $\theta=m$ (for the $+$ sign)
and $\theta=2\pi-m$ (for the $-$ sign). These functions are normalized
to one according to the relation $\int d\mu(u) \mathbb I^\pm_m(u)=1$.
If the fields $\Phi_\pm(x)$ are supposed to be real, 
the matrices $\widehat{\Phi}^j$ are hermitian, and then $\alpha$ and $\beta$ are complex conjugate functions.
As a result, we obtain the general solution for the non-commutative free field written in the fuzzy space 
representation. 

Using the mapping $\mathfrak m$, one can reformulate this solution in terms of functions on $\mathbb E^3$. 
The  components of $\Phi$ are given by:
\begin{eqnarray}
\Phi_\pm(x) \; = \; 
\frac{\ell_P^2}{16\pi}\frac{\sin^2\frac{m}{2}}{\cos\frac{m}{2}}\int_{B_{\ell_P}} \!\!\! d^3p \,
\delta(p-\mu) \; \left( {\alpha}_\pm(p) 
e^{{i}p\cdot x} 
\; + \; {\beta}_\pm(p) e^{-{i}p\cdot x}\right)
\end{eqnarray}
where $B_{\ell_P}$ is the Planck ball, ${\alpha}_\pm(p)=\alpha(u(p))$ where $u(p)$ is the inverse of $p(u)$ when 
$u$ is restricted to the sets $U_\pm$; a similar definition holds for $\beta_\pm$.
We recover the 
usual solution for classical free scalar fields with  the fact that the mass has an upper limit given by $\ell_P^{-1}$. 
Therefore, the Planck mass appears to be a natural UV cut-off.

\subsection{Particles evolving in the fuzzy space}
Important discrepancies between classical and fuzzy dynamics appear when one considers non-linear interactions.
In the case where we study the dynamics of a sole field $\phi$, we may introduce self-interactions. 
However, even in the standard classical commutative space $\mathbb E^3$, 
classical solutions of self-interacting field cannot be written in a closed form in general; and then one cannot expect 
to find explicit solutions for the self-interacting field evolving in the fuzzy background. 
Faced with such technical difficulties (which we postpone for future investigations), 
we will consider simpler models. We will perform symmetry reductions in order 
that the field $\phi$ depends only on one coordinate out of the three.
We will interpret this model as describing one particle evolving in (Euclidean) fuzzy space-time.

\subsection{Reduction to one dimension}
Let us define the algebra $C^{1D}$ of symmetry reduced fields and its different 
representations: the group algebra, the matrix and the continuous formulations. 

First of all, $C^{1D}$ can be identified to the convolution algebra $C(U(1))^*$ of $U(1)$ distributions.
In particular, a function $\varphi \in C^{1D}$ is a function of $\theta \in [0,2\pi]$.
The matrix representation reduces to the Fourier representation of $C(U(1))^*$:
\begin{eqnarray}
{\cal F}^{1D} \; : \; C(U(1))^* \; \longrightarrow \; \textrm{Diag}_\infty(\mathbb C) \;\;,\;\;\;\;
\varphi \; \longmapsto \; \widehat{\Phi} \;\; \textrm{with} 
\;\;\widehat{\Phi}^a_a \; \equiv \; \varphi_a \; = \; \langle \varphi , e^{ia\theta}\rangle \;{\rm~~~~~~~}
\end{eqnarray}
where $\langle,\rangle$ is the duality bracket between $U(1)$ distributions and $U(1)$ functions
and $Diag_\infty(\mathbb C)$ is the algebra of infinite dimensional diagonal complex matrices.
This identity reduces  to the following more concrete relation when $\varphi$ is supposed to be a 
function:
\begin{eqnarray}
\varphi_a \; = \; \frac{1}{2\pi}\int_{0}^{2\pi}\!\!{d\theta} \; {\varphi}(\theta) e^{ia\theta} \;.
\end{eqnarray}
The algebraic structure of $\textrm{Diag}_\infty(\mathbb C)$ is induced from the convolution product $\circ$
and is simply given by the commutative discrete pointwise product:
\begin{eqnarray}
\forall \; \varphi,\varphi' \in C(U(1))^* \;\;\;\;
(\varphi \circ \varphi')_a \; = \; \varphi_a \; \varphi'_a \;.
\end{eqnarray}
Let us now construct the mapping between the convolution algebra $C(U(1))$ and 
the algebra ${C}_{\ell_P}(\mathbb E^1)$ which has to be understood for the moment as 
the one-dimensional analogue of ${C}_{\ell_P}(\mathbb E^3)$. We proceed in the same way as
in the full theory: 
\begin{enumerate}
\item first, we cut $U(1)\equiv[0,2\pi]$ in two parts, $U_+\equiv]-\frac{\pi}{2},\frac{\pi}{2}[$ and 
$U_-\equiv]\frac{\pi}{2},\frac{3\pi}{2}[$ where the symbol $\equiv$ means equal modulo $2\pi$; 
\item then, we construct two bijections between $U_\pm$ and 
$B_{\ell_P}^{1D}\equiv]-\ell_P^{-1};\ell_P^{-1}[$ by assigning to each $\theta \in U_\pm$ a momentum 
$P(\theta)=\ell_P^{-1}\sin\theta$; 
\item the third step consists in associating to any function $\varphi \in C(U(1))$
a pair of functions $\varphi_\pm \in C(U_\pm)$, and a pair of functions $\psi_\pm \in C(\mathbb R)$ induced by 
the previous bijections as follows
\begin{eqnarray}
\mathfrak a^{1D}_\pm(\phi_\pm)(p)\equiv 
\psi_{\pm}(p)  =  \int\frac{d\theta}{2\pi} \, \delta(p-\ell_P^{-1}\sin\theta) \, \varphi_\pm(\theta) \;;
\end{eqnarray}
\item finally, we make use of the standard one dimensional Fourier transform $\mathfrak F^{1D}$ to 
construct the mapping $\mathfrak m^{1D}=\mathfrak m_+^{1D} \oplus  \mathfrak m_-^{1D}:C(U(1))\rightarrow 
C_{\ell_P}(\mathbb E^1)$ where the components $\mathfrak m_\pm^{1D} = {\mathfrak F}^{1D}\circ \mathfrak a_\pm^{1D}$ 
are given by:
\begin{eqnarray}
\mathfrak m_\pm^{1D}(\varphi_\pm)(t)  \equiv  
\Phi_\pm(t) \, = \, \int_0^{2\pi} \frac{d\theta}{2\pi} \, \varphi_\pm(\theta) \, \exp(iP(\theta)t)\,.
\end{eqnarray}
The space $C_{\ell_P}(\mathbb E^1)$ is the image of $C(U(1))$ by $\mathfrak m$ and therefore
is defined by $\widetilde{C}(U_+) \oplus \widetilde{C}(U_-)$ where $\widetilde{C}(U_\pm)$ are the image
by $\mathfrak F^{1D}$ of ${C}(U_\pm)$. As in the full theory, this construction can be extended to the algebra
$C(U(1))^*$ of distributions.
\end{enumerate}
The link between the discrete and the continuous representations of $C^{1D}$ is given by:
\begin{eqnarray}\label{relreduced}
\Phi_\pm(t) \; = \; \sum_a \varphi_a \, K^a_\pm(t) 
\end{eqnarray}
where the functions $K_\pm^a(t)$ are defined by the integrals
\begin{eqnarray}
K^a_\pm(t)  \equiv  \int_{U_\pm}\frac{d\theta}{2\pi} e^{-ia\theta+iP(\theta)t} 
 =  (\pm 1)^a \int_0^{\frac{\pi}{2}} \frac{d\theta}{\pi} \, \cos(a\theta \mp \frac{t}{\ell_P}\sin \theta) \;.
\end{eqnarray}
As in the general case, the relation (\ref{relreduced}) is invertible. The integral defining $K_\pm$
is a simplified version of the general formula (\ref{def de K}) and one can viewed these functions as the
components of the element $K=K_+\oplus K_- \in \textrm{Diag}_\infty(\mathbb C)\otimes C_{\ell_P}(\mathbb E^1)$.
Furthermore, $K^a=K^a_+\oplus K^a_-$ is the image by $\mathfrak m^{1D}$ of the (discrete) plane waves
$\exp(-ia\theta)$. As a final remark, let us underline that $K_+$ and $K_-$ are closely related
by the property $K_-^a(-t)=(-1)^aK_+(t)$. This implies that the functions $\Phi_\pm$ are also closely related:
if we assume for instance that $\varphi_{2n+1}=0$ for any $n \in \mathbb Z$ then $\Phi_-(-t)=\Phi_+(t)$;
if we assume on the contrary that $\varphi_{2n}=0$ for any $n \in \mathbb Z$ then $\Phi_-(-t)=-\Phi_+(t)$.
Such a property will have physical consequences as we will see in the sequel.

Let us give some physical interpretation of the formula (\ref{relreduced}). One can view it as a way to extend
$\varphi_a$, considered as a function on $\mathbb Z$, into the whole real line $\mathbb R$. In that sense,
this formula is a link between the discrete quantum description of a field and a continuous classical description.
One sees that any microscopic time $a$ contributes (positively or negatively) to the definition of a macroscopic 
time $t$
with an amplitude precisely given by $K^a_\pm(t)$. At the classical limit 
$\ell_P \rightarrow 0$, $K^a_\pm(t)$ are maximal for values of the time $t=\pm\ell_P a$.
In other words, the more 
the microscopic time $a\ell_P$ is close to the macroscopic time $t$, the more the amplitude $K^a_\pm(t)$
is important. 

Concerning the reduced $\star$-product, it is completely determined by the algebra of the functions $K^a$
viewed as elements of $C_{\ell_P}(\mathbb E^1)$ and a straightforward calculation leads to the following product
between $K^a$ type functions:
\begin{eqnarray}
K^a \star K^b \; \equiv \; \mathfrak m^{1D}(\exp(-ia\theta) \circ \exp(-ib\theta)) \,  = \, \delta^{ab} K_a \,.
\end{eqnarray}
This result clearly illustrates the non-locality of the $\star$-product.

Before going to the dynamics, let us give the expression of the derivative operator $\partial_t$. 
As for the general case, $\partial_t$ is a finite difference operator whose action on 
$\textrm{Diag}_\infty(\mathbb C)$ is given, as expected, by the following formula:
\begin{eqnarray}
(\partial_t \varphi)_a \; = \; \frac{1}{2\ell_P}\left(\varphi_{a+1} - \varphi_{a-1} \right) \;.
\end{eqnarray}
This expression is highly simplified compared to the more general one introduced in the previous section.
However, we still have the property that $\partial_t$ is in fact a second order operator for it relates $a+1$
and $a-1$. A important consequence would be that the  dynamics (of the free field) will decouple the odd components
$\varphi_{2n}$ and the even components $\varphi_{2n+1}$ of the discrete field. Then, we will have two independent 
dynamics which could be interpreted as two independent particles evolving in the fuzzy space.
In particular, one could associate the continuous fields $\Phi(t)^{odd}$ and $\Phi(t)^{even}$ respectively 
to the families $(\varphi_{2n})$ and $(\varphi_{2n+1})$. It is clear that $\Phi(t)^{odd}$ and
$\Phi(t)^{even}$ are completely independent of one another and, 
using the basic properties of $K_\pm$, we find that the $\pm$
components  of each field are related by:
\begin{eqnarray}
\Phi_-^{odd}(-t) = \Phi_+^{odd}(t) \;\;\;\textrm{and}\;\;\;
\Phi_-^{even}(-t) = -\Phi_+^{even}(t)\;.
\end{eqnarray}
Thus, $\Phi_+$ and $\Phi_-$ fundamentally describe two ``mirror" particles.

\subsubsection{Dynamics of a particle: linear vs. non linear}
We have now assembled all the ingredients for studying the behavior of the one dimensional field $\varphi$. 
When written in the continuous representation, its dynamics is governed by an action of type 
(\ref{action}), but only one-dimensional, with no external field $J$,
and the potential is supposed to be monomial, that is, of the form 
$V(\Phi)=\varepsilon/(\alpha+1) \Phi^{\star (\alpha+1)}$,
with $\alpha+1$ a non-null integer. 
The equations of motions are given by:
\begin{eqnarray}
\Delta \Phi \, + \, \varepsilon  \Phi^{\star \alpha} \; = \; 0 
\end{eqnarray}
where $\Delta=\partial_t^2$.
In the fuzzy space formulation, these equations read:
\begin{eqnarray}\label{dynamics_particle}
\frac{\varphi_{a+2} \; - \; 2\varphi_a\; + \; \varphi_{a-2}}{4\ell_P^2} \; = \;
-\varepsilon \varphi_a^{\alpha} \;.
\end{eqnarray}
As was previously emphasized, 
we  note that these equations do not couple  odd and even integers $a$. For simplicity,
we will consider only even spins, i.e. we assume that $\varphi_{2n+1}=0$ for all integer $n$.

The linear case has already been studied in the previous Section.
For present purposes, we consider the dynamics (\ref{dynamics_particle}) with $\alpha \geq 2$ and 
we look for perturbative solutions in the parameter $\varepsilon$. The corresponding classical solution $\Phi_c$ 
reads at the first order
\begin{eqnarray}\label{classical_solution}
\Phi_c(t) \; = \; vt \; - \; \varepsilon \frac{v^\alpha \; t^{\alpha +2}}{(\alpha+1)(\alpha+2)} \; + \; 
{\cal O}(\varepsilon^2)  
\end{eqnarray}
where we assume for simplicity that $\Phi_c(0)=0$ and $\Phi_c'(0)=v$.

The perturbative expansion of the fuzzy solution is obtained using the same techniques. We look for solutions of
the type $\varphi_a=\lambda a+\varepsilon \eta_a$ where $a=2k$ by assumption,
$\lambda$ is a real number and $\eta$ must satisfy the following relation:
\begin{eqnarray*}
\eta_{2k}  -  \eta_{2k-2} & = & -\frac{\ell_P^2}{4} \lambda^\alpha \sum_{n=1}^{k-1} (2n)^\alpha \; = \; 
 -\frac{\ell_P^2}{4} (2\lambda)^\alpha [\frac{(k-1)^{\alpha +1}}{\alpha +1} + \frac{(k-1)^\alpha}{2} \\
 & & + \frac{\alpha(k-1)^{\alpha -1}}{12} - \frac{\alpha(\alpha-1)(\alpha-2)}{720}(k-1)^{\alpha-3}\\
& & + \frac{\alpha(\alpha -1)(\alpha-2)(\alpha-3)(\alpha-4)}{30240}(k-1)^{\alpha-5} + \cdots]
\end{eqnarray*}
The solution is in general complicated.
To be explicit, we will consider the case $\alpha=2$. The formula simplifies considerably 
and, after some straightforward calculations, one can show that:
\begin{eqnarray}
\eta_{2k} \; = \; -\frac{\ell_P^2\lambda^2}{12}k^2(k-1)(k+1) \; = \;  
-\frac{\ell_P^2\lambda^2}{12} (k^4-k^2)\,.
\end{eqnarray}
In order to compute the $C_{\ell_P}(\mathbb E^1)$ representation of this solution,
one uses the following relations for any integer $n$
\begin{eqnarray}
S_\pm^{(n)}(t) \; \equiv \; 
\sum_{k=-\infty}^{+\infty} k^n K_\pm^{2k}(t) \; = \; \frac{1}{2(2i)^n} \frac{d^n}{d\theta^n}\exp(iP(\theta)t)
\vert_{\frac{1\mp 1}{2}\pi} \;.
\end{eqnarray}
Applying this formula for $n=1,2$ and $4$
\begin{eqnarray} \label{identity for S}
S_\pm^{(1)}(t)=\pm \ell_P^{-1}t ,\;\;\;
S_\pm^{(2)}(t)=2\ell_P^{-2}t^2 ,\;\;\; 
S_\pm^{(4)}(t)= 2(4\ell_P^{-4}t^4+\ell_P^{-2}t^2)
\end{eqnarray}
one shows, after some simple calculations, 
that $\Phi_+$ and $\Phi_-$  are simply related by $\Phi_+(t)=\Phi_-(-t)$ and $\Phi_+$ is given by:
\begin{eqnarray}
\Phi_+(t) \, = \, 2\lambda \ell_P^{-1}t \, - \, \varepsilon \frac{\ell_P^2 \lambda^2}{6}
(2\ell_P^{-4}t^4 + \ell_P^{-2}t^2) \; + \; {\cal O}(\varepsilon^2)\,.
\end{eqnarray}
To compare it with the classical solution $\Phi_c$ computed above
(\ref{classical_solution}), we impose the same initial conditions which leads to 
$\lambda=v\ell_P/2$ and then the solution reads:
\begin{eqnarray}
\Phi_+(t) \; = \; vt - \varepsilon \frac{v^2t^4}{12} \, - \, \varepsilon \frac{\ell_P^2 v^2 t^2}{24} \, + \, {\cal O}(\varepsilon^2) \;.
\end{eqnarray}
Let us interpret the solution.
First, let us underline that $\Phi_+$ and $\Phi_-$ are related by $\Phi_+(t)=\Phi_-(-t)$: 
thus, it seems that $\Phi_-$ corresponds to a particle
evolving backwards compared to $\Phi_+$. In that sense, the couple $\Phi_\pm$ behaves like a particle 
and a ``mirror" particle: the presence of the mirror particle is due to quantum gravity effects.
Second, we remark that the solution for $\Phi_+$  differs from its classical counterpart at least 
order by order in the parameter $\varepsilon$.
At the no-gravity limit $\ell_P \rightarrow 0$, $\Phi_+$ tends to the classical solution
(\ref{classical_solution}). 
Therefore, we can interpret these discrepancies as an illustration of quantum gravity effects
on the dynamics of a field. 

\subsubsection{Background independent motion}
We finish this example with the question concerning the physical content of this solution.
For the reasons given in the previous section, we concentrate only on the component $\Phi_+$.
Can one interpret $\Phi_+(t)$ as the position $q(t)$ of a particle evolving in the fuzzy space? 
If the answer is positive,
it is quite confusing because the position should be discrete valued whereas $\Phi_+$ takes value in the
whole real line a priori. In fact, we would like to interpret $\Phi_+(t)=Q(t) \in \mathbb R$ as the extension in 
the whole real 
line of a discrete position $q(t) \in \mathbb Z$. More precisely, we suppose that the space where the
particle evolves is one-dimensional and discrete, and then its motion should be characterized by a $\mathbb Z$-valued
function $q(t)$. If we restore the discreteness of the time variable, then the motion of the particle should, 
in fact, be characterized by a set of ordered integers $\{q(2k\ell_P),k\in \mathbb Z\}$.
To make this description more concrete, we make use of the identity satisfied by $S_+^{(1)}$ 
(\ref{identity for S})  which implies that:
\begin{eqnarray}
 Q(t)\;=\; \sum_{k=-\infty}^{+\infty} (2\ell_P k) \, K^{2k}_+(Q(t)) \;.
\end{eqnarray}
This identity makes clear that $Q(t)$ can be interpreted as a kind of continuation in the whole real line
of a set of discrete positions and $K^{2k}_+(Q(t))$ gives
the (positive or negative) weight of the discrete point $2\ell_P k$ in the evaluation of the continuous point $Q(t)$. 
Therefore, one can associate an amplitude ${\cal P}(k \vert \tau)$ to the particle 
when it is at the discrete position $Q=2\ell_P k$ and at the discrete time $t=2 \ell_P \tau$ (in Planck units) 
in the fuzzy space. This amplitude is given by:
\begin{eqnarray}\label{ampliP}
{\cal P}(k\vert\tau) \; = \; 
\frac{K^{2k}_+(Q(2\ell_P \tau))}{\sum_{j=-\infty}^{+\infty}K^{2j}_+(Q(2\ell_P \tau))} \; = \; 
K^{2k}_+(Q(2\ell_P \tau)) 
\end{eqnarray}
because the normalisation factor equals one. 
These amplitudes cannot really be interpreted as statistical weights because they can be positive or negative.
Nevertheless, they contain all the information of the dynamics of the particle in the sense that one can 
reconstruct the dynamic from these data. 
Therefore, we obtain a background independent description of the dynamics
of the particle that can be a priori anywhere at any time: its position $2k\ell_P$ at a given time 
$2\tau \ell_P$ is characterized
by the amplitude previously defined. 
Furthermore, the amplitude is maximum around the classical trajectory, i.e. when
$Q(2\ell_P \tau)=2\ell_P k$, and gives
back the classical trajectory at the classical limit defined by $k,\tau \rightarrow \infty$, $\ell_P \rightarrow 0$ with
the products $k\ell_P$ and $\tau\ell_P$ respectively fixed to the values $t$ (classical time) and $Q$ (classical position).

\section{Discussion}
In this article, we have tackled the question of motion in Quantum Gravity.
As we have already emphasized, it is certainly too early to discuss this question in detail. However 
it is at least possible to raise some preliminary problems that one needs to resolve if one aims
at understanding what motion means at the Planck scale. Among the most fundamental problems are
the questions of the deep structure of space-time and those of the description of matter fields in
Quantum Gravity. Loop Quantum Gravity proposes a very clear answer to these questions (even if the 
specific viewpoint adopted here has warranted extensive discussion). 

For this reason, we think that Loop Quantum Gravity presents a useful framework for discussing the question of
motion at very short distances. We started with a very brief review of Loop Quantum Gravity, insisting
on the kinematical aspects: the description of the kinematical states in terms of spin-networks
and the computation of the spectrum of the so-called area and volume operators. We explained in what sense
space appears discrete in Loop Quantum Gravity. We finished by mentioning the fundamental problem of the dynamics
that, so far, no one knows how to solve, namely, the remaining Hamiltonian constraint. Nonetheless, 
different promising strategies have been developed to solve this issue. For the moment, this relative failure
prevents us from discussing the question of the motion which is intimately linked to the question of the dynamics.

As a consequence, in a second part, we presented a toy model where the dynamics is very well-understood: three dimensional
Euclidean quantum gravity with no cosmological constant. This model is exactly solvable and shares several characteristics
with Loop Quantum Gravity, including the discreteness of space. Furthermore, the coupling to a matter field is very well understood and leads
to a description of scalar fields in terms of complex matrices evolving in a non-commutative fuzzy geometry. Therefore, the question of
motion at the Planck scale reduces in that case to the resolution of finite difference equations involving matrix coefficients which are
the quantum analogue of the equations of motions. We propose a solution of these equations in simple examples: the free field and the one-dimensional
field. These examples are nice illustrations of what could represent motion more generally in Quantum Gravity.

It is nonetheless clear that we are far from a precise description of motion in full Loop Quantum Gravity. We hope that
the examples we have developed in this paper will shed light on this problem.



\end{document}